\documentclass[a4paper,twocolumn]{esapub} 
\usepackage{natbib}
\usepackage{epsfig}

\title{THE CYGNUS X REGION: A NUCLEOSYNTHESIS LABORATORY FOR INTEGRAL}
\author[1]{J. Kn\"odlseder}
\author[1,2]{M. Cervi\~no}
\author[2]{D. Schaerer}
\author[1]{P. von Ballmoos}
\author[3]{G. Meynet}
\affil[1]{Centre d'Etude Spatiale des Rayonnements, B.P. 4346, 31028 
Toulouse Cedex 4, FRANCE}
\affil[2]{Observatoire Midi-Pyr\'en\'ees, 14, avenue Edouard Belin, 
31400 Toulouse, FRANCE}
\affil[3]{Observatoire de Gen\`eve, CH-1290 Sauverny, SWITZERLAND}

\newcommand{\flxal}{\mbox{$\Phi_{\rm 1.809 \hspace{0.2em} MeV}$}}
\newcommand{\flxfe}{\mbox{$\Phi_{\rm 1.137 \hspace{0.2em} MeV}$}}
\newcommand{\flxff}{\mbox{$\Phi_{\rm 53 \hspace{0.2em} GHz}$}}
\newcommand{\yal}{\mbox{Y$_{26}^{\rm O7\hspace{0.2em}V}$}}

\newcommand{\HII}{\mbox{H\hspace{0.2em}{\scriptsize II}}}
\newcommand{\CII}{\mbox{$[$C\hspace{0.2em}{\scriptsize II}$]$}}
\newcommand{\al}{\mbox{$^{26}$\hspace{-0.2em}Al}}
\newcommand{\fe}{\mbox{$^{60}$Fe}}
\newcommand{\Msol}{\mbox{$M_{\sun}$}}

\newcommand{\pcmq}{\mbox{cm$^{-2}$}}

\newcommand{\psec}{\mbox{s$^{-1}$}}

\newcommand{\funit}{\mbox{ph \pcmq \psec}}

\def\MeV{\mbox{Me\hspace{-0.1em}V}}

\def\deg{\ensuremath{^\circ}}
\def\sun{\hbox{$\odot$}}

\begin{document}


\maketitle

\begin{abstract}
The detection of 1.809 MeV gamma-ray line emission from the Cygnus X 
complex by the COMPTEL telescope is one of the 
most convincing proves of massive star nucleosynthesis in our Galaxy.
The Cygnus X complex is an extremely active nearby region, containing 
several star forming regions, OB associations and young open star 
clusters.
It houses some of the most massive stars known in our Galaxy and 
concentrates a large number of Wolf-Rayet stars.
Thermal radio continuum emission and intense \CII\ line emission 
reveals widespread ionisation, and at least 60 individual \HII\ regions 
have been identified.
In order to understand the 1.809 MeV gamma-ray line emission from the 
Cygnus X complex, and to compare the observations to theoretical 
nucleosynthesis calculations, we modelled the multi-wavelength 
spectrum of the region by means of an evolutionary synthesis model.
Our investigation leads us to the following conclusions:
\begin{enumerate}
\item Stellar wind ejection is the dominant mechanism for the 
      observed \al\ enrichment in the Cygnus X region
\item Cyg OB2 is by far the dominating massive star association in 
      Cygnus X and 1.809 MeV line emission from \al\ produced in 
      this association should be detectable by the spectrometer SPI on 
      {\em INTEGRAL}
\item There is only low supernova activity in the Cygnus X complex and 
      the \fe\ lines should be below {\em INTEGRAL's} detection 
      sensitivity 
\end{enumerate}

The detectability of an individual massive star cluster (Cyg OB2) by 
{\em INTEGRAL} provides a fantastic opportunity for future nucleosynthesis 
studies using gamma-ray line spectroscopy.
We will explore the scientific potential of such observations, and 
provide estimates for gamma-ray line intensity distributions based on 
the massive star census of the Cygnus X region.
\end{abstract}

\section{Introduction}

Cygnus X designates one of the most prominent extended emission 
regions in galactic radio continuum surveys, located between 
$l=70\deg - 90\deg$ and $b=\pm5\deg$.
At high angular resolution, the emission breaks into hundreds of 
continuum sources \citep{wendker84}, most optically invisible or 
heavily reddened, largely owing to foreground obscuration by the 
Cygnus Rift at a distance of 700 pc.

1.809 \MeV\ gamma-ray line emission from the Cygnus X region has been 
first reported by \citet{delrio96} based on the analysis of 2 years 
of COMPTEL observations.
They observed a diffuse extended emission feature with a total 1.809 
\MeV\ line flux of $7\,10^{-5}$ \funit.
To understand the origin of this feature, \citet{delrio96} 
modelled \al\ nucleosynthesis by adding the expected individual 
contributions from known Wolf-Rayet (WR) stars and supernova 
remnants (SNRs) in this area.
In that way they predict a total 1.809 \MeV\ flux of $4.2\,10^{-5}$ 
\funit\ from which $\sim85\%$ is attributed to WR stars.
A close inspection of their list of candidate sources reveals, however, 
that this flux estimate is dominated by a few nearby ($<1$ kpc) objects.
Since distance estimates for individual WR stars and SNRs are quite 
uncertain, it is questionable how accurate such a candidate source 
census can be.

In this work we present an alternative approach to model \al\ production in 
Cygnus which is based on a census of massive star populations in 
the field (such as OB associations and young open clusters).
The basic parameters, like distance and age, are determined much 
more reliable for stellar ensembles than for individual objects, 
hence we believe that our model is more accurate in the prediction of 
massive star nucleosynthesis.
On the other hand, massive stars that are not situated in associations 
are not accounted for by our model, so strictly speaking we only 
obtain a lower limit.
Globally, however, most massive stars are located in associations 
\citep{garmany94}, hence the underestimation should at most amount to 
$20\%$ (we will show later that it probably is even smaller).
Additionally, our approach will be much less sensitive to 
completeness limits than the model of \citet{delrio96}, since the 
population parameters can be estimated reasonably well without knowing 
all members of the population.
In particular, we also estimate nucleosynthetic contributions from 
stars that already faded away by means of an evolutionary synthesis 
code (Cervi\~no et al., 2000; see also Cervi\~no et al., these 
proceedings).
Although \citet{delrio96} also try to estimate this contribution 
from known radio supernova remnants, it is clear that this approach 
probably misses an important fraction of recent \al\ production due 
to incompleteness of SNR catalogues.
In particular old remnants are difficult to detect against the 
galactic background radiation \citep{green91}, but their \al\ still 
contributes to the today 1.809 \MeV\ radiation.

\section{Free-free emission}

The radio continuum emission from the Cygnus X region is primarily 
optically thin thermal radiation that arises from individual \HII\ 
regions and diffuse ionised gas \citep{wendker70}.
Ionised gas is also nicely traced by the far-infrared transition of 
\CII\ at 158 $\mu$m, and the correlation of diffuse \CII\ and thermal 
radio emission in Cygnus X is further evidence for the presence of 
abundant ionised gas in this region \citep{doi97}.

Thermal radio emission has also shown to be important in the 
context of \al.
\citet{knoedl99} have demonstrated that the 1.809 \MeV\ intensity 
distribution follows closely the galactic distribution of free-free 
emission, mapped for example by the DMR instrument aboard {\em COBE} 
at 53 GHz \citep{bennett92}.
In fact, this correlation is one of the strongest pieces of evidence 
that \al\ production arises from massive stars which also power the 
free-free emission by ionisation of the interstellar medium.
To express \al\ production in terms of ionisation, \citet{knoedl99apj} 
proposed to normalise the 1.809 \MeV\ emission on the number of 
ionising photons produced by an representative ionising star (e.g.~an O7V 
star), leading to an average equivalent O7V star yield of
$\yal = (1.0 \pm 0.5)\,10^{-4}$ \Msol\ for the entire Galaxy.

From the COMPTEL 1.809 \MeV\ 7 years all-sky map, \citet{plueschke00} 
derived a total gamma-ray line flux of $\flxal = (7.9 \pm 2.4)\,10^{-5}$ 
\funit\ for the Cygnus region, in reasonable agreement with the earlier 
determination by \citet{delrio96} from only 2 years of data.
Integrating the intensity in the DMR 53 GHz free-free emission map 
over the same region results in a total flux of $\flxff = 5400 \pm 1200$ Jy, 
where the quoted error reflects the uncertainty that arises from the 
background level determination and the subtraction of an underlying 
galactic ridge emission that is probably not associated to the Cygnus 
X complex.
The ratio between both flux measurements is then converted into the 
equivalent yield (in units of \Msol) using
\begin{equation}
 \yal = 7.91\,10^{3} \times \frac{\flxal\ (\funit)}{\flxff\ (\rm Jy)} .
\end{equation}
Thus, for the Cygnus X region we obtain $\yal = (1.2 \pm 0.4)\,10^{-4}$ 
\Msol, compatible with the galactic average value.

\section{Massive star populations}
\label{sec:population}

Our massive star census of the Cygnus region is based on the WEBDA 
catalogue of \citet{mermilliod98} for open clusters and the 
compilations of \citet{garmany92} and \citet{humphreys78} for 
galactic OB associations.
The data have been complemented by information from the SIMBAD 
database, recent analyses of stellar associations by Hipparcos 
\citep{deZeeuw99}, and a recent investigation of the stellar 
population of Cyg OB2 \citep{knoedl00}.
For the investigated field $60\deg<l<110\deg$ and $|b|<15\deg$, our 
final database contains 10 associations and 21 young open clusters, 
where young means an age of less than 50 Myrs, corresponding to the 
lifetime of a 8 \Msol\ star (i.e. the lowest initial mass that is 
believed to lead to a type II supernova).
We find a total of 182 O stars within the OB associations, while the 
clusters only contain 19 O stars.

\begin{figure}
  \epsfxsize=8cm \epsfclipon
  \epsfbox{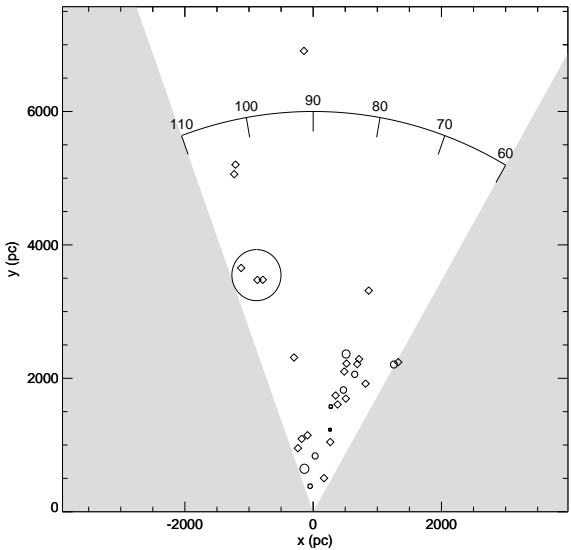}
  \caption{\label{fig:spiral}
    Distribution of OB association (circles) and young open clusters (diamonds) 
    in the  Cepheus-Cygnus region projected onto the galactic plane
    (the positive x-axis points towards the galactic centre).
    The grey-shaded area delineates the boundaries of the investigated
    field, the labelled arc indicates galactic longitudes.
  }
\end{figure}

To estimate the fraction of massive stars in Cygnus that is included in 
our association census, we extracted from the SIMBAD database all stars 
in the field that are classified as type O.
Following the discussion of \citet{garmany82} this sample should be 
complete to $\sim2-3$ kpc, corresponding to the maximum distances of 
massive star associations found towards Cygnus X (see Fig.~\ref{fig:spiral}).
An exception might be the heavily reddened Cyg OB2 association that 
contains about 120 O stars \citep{knoedl00} that are only partially 
identified in SIMBAD.
Correcting for this underestimation, we find a total of 223 classified O 
stars in the field from which 185 (or 83\%) are found in one of the 
associations or clusters of our database.
The remaining field O stars are mainly found towards Cep OB2 and 
north-east of Cyg OB3 showing a wide spread in distances between $1-5$ 
kpc.
Thus we believe that our database present a fairly complete 
description of the massive star census in Cygnus, and that the small 
($17\%$) fraction of field stars and their large distances will not lead 
to a severe underestimation of the nucleosynthetic production in this 
area.

For each of the associations or clusters in our database we estimated 
the distance by the method of spectroscopic and photometric parallaxes.
H-R diagrams were built from spectroscopic and photometric data from 
which cluster ages have been determined by isochrone fitting.
Particular care has been attributed to the estimation of age and 
distance uncertainties since they directly translate into uncertainties 
of our model predictions \citep{cervino00}.
We also derived initial mass functions from the data to estimate the 
initial richness of the populations (a detailed description of the 
methods will be presented in a forthcoming paper).

To illustrate the spatial distribution of the investigated 
populations, we used our distance estimates to project the associations 
and clusters onto the galactic plane (see Fig.~\ref{fig:spiral}).
The most prominent feature in this representation is an elongated 
structure running from the solar vicinity towards $l\sim70\deg$ to a 
distance of $\sim3$ kpc, known as the local spiral arm.
From the 201 O stars in our database 200 lie within associations and 
clusters of the local spiral arm, illustrating the extremely young age 
of this structure.
$60\%$ of the O stars lie within a single association, Cyg OB2, situated at 
a distance of $\sim1.6$ kpc.
The extreme richness of this association, as revealed by near-infrared 
observations, has recently led one of us \citep{knoedl00} to identify 
Cyg OB2 as a young globular galactic cluster.
From isochrone fitting we estimate the age of this cluster to 
$2.5\pm1.0$ Myrs.

\section{Model results}

\begin{table}[h!]
  \begin{center}
    \caption{Model predictions versus observations. Gamma-ray fluxes are 
    given in units of $10^{-5}$ \funit.
    }\vspace{1em}
    \renewcommand{\arraystretch}{1.2}
    \begin{tabular}[h]{lcc}
      \hline
      & predicted 
      & observed \\
      \hline
      \flxal\                 & $1.8 - 5.4$   & $7.9 \pm 2.4$ \\
      \flxff\ (Jy)            & $2900 - 9300$ & $5400 \pm 1200$ \\
      \yal\ ($10^{-4}$ \Msol) & $0.2 - 1.0$   & $1.2 \pm 0.4$ \\
      \flxfe\                 & $0.1 - 0.4$   & - \\
      \hline \\
    \end{tabular}
    \label{tab:results}
  \end{center}
\end{table}

Using our evolutionary synthesis model (Cervi\~no et al., 2000; see also 
Cervi\~no et al., these proceedings), we estimated the integrated
gamma-ray line and free-free radiation fluxes for the entire Cygnus 
region.
In addition to \al\ we also included \fe\ nucleosynthesis in our 
model to predict also the gamma-ray flux from the radioactive decay 
of this longlived isotope.
Initial mass function slopes between $\Gamma = -1.1$ and $-1.35$ have 
been explored which covers the range of values that we found for the 
associations in our database.
Our model predictions are summarised in Table \ref{tab:results}.
For each quantity of interest we predict a range of possible values 
that includes uncertainties in distance, age, richness, evolution, and 
IMF slope for each association.
We deliberately did not include uncertainties related to theoretical
nucleosynthesis yields, since we consider our model as a hypothesis that 
we want to test against observations of gamma-ray line emission.
In total we predict a 1.809 \MeV\ gamma-ray line flux between 
$(1.8-5.4)\,10^{-5}$ \funit\ where the lower value corresponds to the 
lower flux limit for an IMF slope of $\Gamma=-1.35$ while the upper value 
is the upper flux limit obtained for a flat slope of $\Gamma=-1.1$.
On average, about 80\% of the 1.809 \MeV\ emission in our model 
originates from \al\ ejected by stellar winds while only 20\% comes from 
core-collapse supernovae.
Also, OB associations play a dominant role since they contribute more 
than 90\% of the 1.809 \MeV\ emission in Cygnus.

The most prominent source of \al\ is probably Cyg OB2 for 
which we predict a median 1.809 \MeV\ flux of $8.5\,10^{-6}$ \funit, 
at the detection limit of current COMPTEL observations and 
above the sensitivity limit of SPI for a two-weeks observation.
The flux uncertainty for Cyg OB2 is quite large, ranging from
$1.1\,10^{-6}$ to $2.3\,10^{-5}$ \funit, which mainly reflects our 
adopted age uncertainty of $2.5\pm1.0$ Myrs.
At this age the most massive stars of the association start to evolve 
off the main sequence which generally means a considerable increase of 
the mass-loss rate, either in a LBV or a WR phase, leading to an
enrichment of the interstellar medium with freshly produced \al.
Consequently, a small variation in the adopted age at this stage leads 
to a quite important variation in the number of mass-losing stars (LBV 
and WR), which translates directly in a large variation in the
interstellar \al\ enrichment.
The existence of a few WR stars in Cyg OB2 indicates that, if 
formed coevally, the association should not be younger than $\sim2$ 
Myrs.
On the other hand, Cyg OB2 can not be much older than 3-4 Myrs 
since the absence of supernova remnants within the cluster boundary 
indicates that all stars are still alive \citep{wendker91}.
However, star formation in Cyg OB2 was probably not strictly coeval
\citep{massey91}, hence the indicated age uncertainty merely reflects 
the observed age-spread in this cluster.
In this case, our median flux value corresponds to the age averaged 
1.809 \MeV\ intensity.

\begin{figure}
  \epsfxsize=8cm \epsfclipon
  \epsfbox{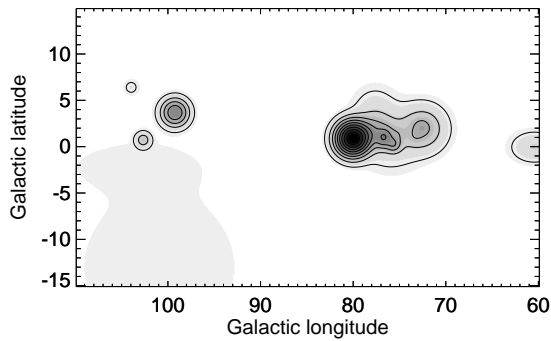}
  \caption{\label{fig:intensity}
    Predicted median 1.809 \MeV\ intensity distribution.
  }
\end{figure}

To illustrate the spatial distribution of 1.809 \MeV\ emission 
predicted by our model we generated an intensity map based on the 
median flux values that we distributed within the association boundaries 
using Gaussian shaped density profiles 
(cf.~Fig.~\ref{fig:intensity}).
This figure may be compared to Fig.~1 of Pl\"uschke et al. 
(these proceedings) which shows the COMPTEL 1.809 \MeV\ maximum 
entropy image of the Cygnus X region.
The emission maximum in the COMPTEL map lies only $3\deg$ from the 
centre of the Cyg OB2 cluster, which is certainly not a proof, but a 
strong indication that Cyg OB2 is indeed a prominent source of \al\ 
(note that the angular resolution of COMPTEL at 1.809 \MeV\ amounts to 
$\sim4\deg$ FWHM).
Additional contributions in this area come from the OB associations 
Cyg OB1, Cyg OB3 and Cyg OB9 which are situated west of Cyg OB2.

However, the predicted 1.809 \MeV\ flux from Cygnus is only marginally 
compatible with the observed flux, a problem that has already been 
encountered in the earlier work of \citet{delrio96}.
We possibly underestimated the massive star population in Cygnus X 
since we did not included stars lying outside associations, yet our discussion 
in Sect.~\ref{sec:population} suggests that our census is fairly complete.
In particular, our model reproduces the observed 53 GHz free-free 
emission reasonably well, and an increase in the number of massive 
stars would quickly lead to an overproduction of ionising photons in 
the Cygnus region.
Additionally, the predicted equivalent yield \yal\ falls also below 
the observed value, indicating that the problem is more likely related 
to an underestimation of the \al\ yield per massive star.
The origin of the low \yal\ values could lie in erroneous 
age determinations, since \yal\ depends strongly on the evolutionary 
state of the associations.
Indeed, some of the OB associations (such as Vul OB1, Cyg OB7, Cyg OB8, 
Cyg OB9 and Cep OB1) show only poorly defined main sequences which 
makes it difficult to estimate their actual age.
It is even questionable if these associations are physically related
\citep{deZeeuw99} which questions the validity of treating their \al\ 
production by means of an evolutionary synthesis code.
However, the number of O stars situated in these associations is 
fairly small (about 17\% of the total number), and it is 
unlikely that they alone are responsible for the discrepancy.
Finally, there remains the possibility that current nucleosynthesis 
models underestimate \al\ production in massive mass-losing stars.
Indeed, a comparison of the observed galactic \al\ mass with a simple 
model for galactic \al\ enrichment also indicates a possible 
underestimation of the global galactic \al\ yield by $30-50\%$ 
\citep{knoedl99apj}.
Recent calculations of \al\ nucleosynthesis in rotating massive stars 
suggest that rotation could considerably enhance \al\ production 
\citep{meynet99}, providing an interesting mechanism that could 
bring the theory in better agreement with the observations.

The predicted 1.137 \MeV\ flux arising from the decay of \fe\ is 
rather small, and a detection of this emission by SPI seems highly 
improbable.
The low flux is a result of the low supernova activity in the Cygnus X 
region, which is the dominant production channel for this radioisotope.
On the other hand, the observation of large 1.809 \MeV\ fluxes in 
conjunction with low supernova activity is a remarkable situation.
Although we have no observational evidence that allows us to generalise 
such a correlation for the entire Galaxy, this indicates
that supernovae may play a less important role for galactic \al\ 
production than previously thought.


\end{document}